\documentclass[sigconf,nonacm]{acmart}
\renewcommand\footnotetextcopyrightpermission[1]{} 
\pdfoutput=1 
\acmPrice{}
\acmISBN{ }
\acmBooktitle{ }
\usepackage{multirow}
\usepackage{rotating}
\usepackage{booktabs} 
\usepackage{amsmath}
\usepackage{graphicx}
\usepackage{subcaption}
\usepackage[belowskip=-12pt]{caption}
\usepackage{rotating}
\usepackage{tabularx}
\usepackage{wrapfig}
\usepackage{listings}
\usepackage{color, colortbl}
\usepackage{balance} 
\usepackage{lscape}
\usepackage{hyperref}
\usepackage{xspace}
\usepackage{framed}
\usepackage{syntax}
\usepackage{enumitem}
\usepackage{soul}
\usepackage{flushend}
\usepackage[tikz]{bclogo}
\usepackage{makecell}
\definecolor{codegreen}{rgb}{0,0.6,0}
\definecolor{codegray}{rgb}{0.5,0.5,0.5}
\definecolor{codepurple}{rgb}{0.58,0,0.82}
\definecolor{backcolour}{rgb}{0.95,0.95,0.92}
\definecolor{Gray}{gray}{0.1}
\lstdefinestyle{mystyle}{
	backgroundcolor=\color{backcolour},   
	commentstyle=\color{codegreen},
	keywordstyle=\color{magenta},
	numberstyle=\tiny\color{codegray},
	stringstyle=\color{codepurple},
	basicstyle=\scriptsize,
	breakatwhitespace=false,         
	breaklines=true,                 
	captionpos=b,                    
	keepspaces=true,                 
	numbers=left,                    
	numbersep=5pt,                  
	showspaces=false,                
	showstringspaces=false,
	showtabs=false,                  
	tabsize=2
}

\lstdefinelanguage{Pythonna}{%
	language     = python,
	morekeywords = {to_categorical, flow_from_directory,pad_sequences,load_image},
	morecomment=[f][\color{diffstart}]{@@},
	morecomment=[f][\color{diffincl}]{+\ },
	morecomment=[f][\color{diffrem}]{-\ }	
}

\lstdefinestyle{customc}{
	belowcaptionskip=1\baselineskip,
	breaklines=false,
	frame= single,
	breaklines = true,
	xleftmargin=\parindent,
	language= Pythonna,
	showstringspaces=false,
	keywordstyle=\bfseries\color{green!40!black},
	commentstyle=\itshape\color{purple!40!black},
	identifierstyle=\color{blue},
	stringstyle=\color{codegreen},
	backgroundcolor=\color{gray!4}
}

\graphicspath{{./figures/}}

\newcommand{\fignref}[1]{Figure~\ref{#1}}

\newcommand{\secnref}[1]{\S\ref{#1}}

\newcommand{\etal}{{\em et al.}\xspace}

\newcommand{\bart}{\textit{Bart}\xspace}
\newcommand{\bert}{\textit{Bert}\xspace}
\newcommand{\ctrl}{\textit{CTRL}\xspace}
\newcommand{\gpt}{\textit{GPT-2}\xspace}
\newcommand{\roberta}{\textit{RoBERTa}\xspace}
\newcommand{\tf}{\textit{T5}\xspace}
\newcommand{\txl}{\textit{Transformer-XL}\xspace}

\newcommand{\sof}{\textit{Stack Overflow}\xspace}

\newcommand{\tensor}{\textit{Tensorflow}\xspace}

\newcommand{\torch}{\textit{Pytorch}\xspace}

\newcommand{\gh}{\textit{GitHub}\xspace}

\newcounter{rqs}
\stepcounter{rqs}

\newcounter{NumObservations}
\stepcounter{NumObservations}
\definecolor{shadecolor}{rgb}{.9,.9,.9}

\newcommand{\finding}[1]{%
	\begin{mdframed}[
		backgroundcolor= gray!08,
		linewidth=.2pt,
		linecolor = black,
		roundcorner=2pt,
		skipabove=8pt,
		innertopmargin=5pt,
		innerbottommargin=5pt,
		innerrightmargin=4pt,
		innerleftmargin=2pt,
		leftmargin = 2pt,
		rightmargin = 2pt]%
		\noindent{\textbf{Finding \arabic{NumObservations}}: #1} 
	\end{mdframed}%
	\stepcounter{NumObservations}
}

\newcommand*\circled[1]{\tikz[baseline=(char.base)]{
		\node[shape=circle,fill,inner sep=1pt] (char) {\textcolor{white}{#1}};}}

\lstdefinestyle{mystyle}{
	backgroundcolor=\color{backcolour},   
	commentstyle=\color{codegreen},
	keywordstyle=\color{magenta},
	numberstyle=\tiny\color{codegray},
	stringstyle=\color{codepurple},
	basicstyle=\scriptsize,
	breakatwhitespace=false,         
	breaklines=true,                 
	captionpos=b,                    
	keepspaces=true,                 
	numbers=left,                    
	numbersep=5pt,                  
	showspaces=false,                
	showstringspaces=false,
	showtabs=false,                  
	tabsize=2
}
\AtBeginDocument{%
	\providecommand\BibTeX{{%
			\normalfont B\kern-0.5em{\scshape i\kern-0.25em b}\kern-0.8em\TeX}}}

\graphicspath{{./figures/}}

\begin{document}


\title{An Empirical Study on the Bugs Found while Reusing Pre-trained Natural Language Processing Models}

\author{Rangeet Pan}
\email{rangeet@iastate.edu}
\affiliation{%
	\institution{Iowa State University}
	\city{Ames}
	\state{Iowa}
	\country{USA}}
\author{Sumon Biswas}
\email{sumon@iastate.edu}
\affiliation{%
	\institution{Iowa State University}
	\city{Ames}
	\state{Iowa}
	\country{USA}}
\author{Mohna Chakraborty}
\email{mohnac@iastate.edu}
\affiliation{%
	\institution{Iowa State University}
	\city{Ames}
	\state{Iowa}
	\country{USA}}
\author{Breno Dantas Cruz}
\email{bdantas@iastate.edu}
\affiliation{%
	\institution{Iowa State University}
	\city{Ames}
	\state{Iowa}
	\country{USA}}
\author{Hridesh Rajan}
\email{hridesh@iastate.edu}
\affiliation{%
	\institution{Iowa State University}
	\city{Ames}
	\state{Iowa}
	\country{USA}}

\begin{abstract}
In Natural Language Processing (NLP), reusing pre-trained models instead of training from scratch has gained a lot of popularity; however, these models are mostly black-boxes, extremely large, and building a model from scratch often requires significant resources. To ease, models trained with large corpora are made available, and developers reuse them in different problems. In contrast, developers mostly build their models from scratch for traditional deep learning (DL)-related problems. By doing so, they have full control over the choice of algorithms, data-processing, model structure, tuning hyperparameters, etc. Whereas in NLP, due to the reuse of the pre-trained models, NLP developers are limited to very little to no control over such design decisions. They can either apply tuning or transfer learning on pre-trained models to meet their requirements. Also, NLP models and their corresponding datasets are significantly larger than the traditional DL models and require heavy computation. Such reasons often lead to bugs in the system while reusing the pre-trained models. While bugs in traditional DL software have been intensively studied, the nature of extensive reuse and black-box structure motivates us to understand the different types of bugs that occur while reusing NLP models? What are the root causes of those bugs? How do these bugs affect the system?
To answer these questions, we studied the bugs reported while reusing the 11 popular NLP models. We mined 9,214 issues from their respective \textit{GitHub} repositories and identified 984 bugs. We created a taxonomy with the bug types, root causes, and impacts. Our observations led to several key findings, including limited access to model internals resulting in lack of robustness, lack of input validation leading to the propagation of the algorithmic and data bias, high-resource consumption causing more crashes, and memory-out-of-bound errors,  etc. Our observations suggest several bug patterns, which would greatly facilitate further research and development for reducing bugs in the pre-trained models as well as the code that reuses them.
\end{abstract}



\maketitle

\section{Introduction}
\label{sec:intro}
``Hey, Alexa! How is the weather outside?'', ``Hey, Google! Play my favorite
song'', or ``Hey, Siri! Set the alarm at 6 pm?'' these are a few
examples in which Natural Language Processing (NLP) has been used in our daily lives. In each one, human speech is transformed into machine-level representations.
\begin{figure}[t]
	\centering
	\includegraphics[width=0.97\linewidth]{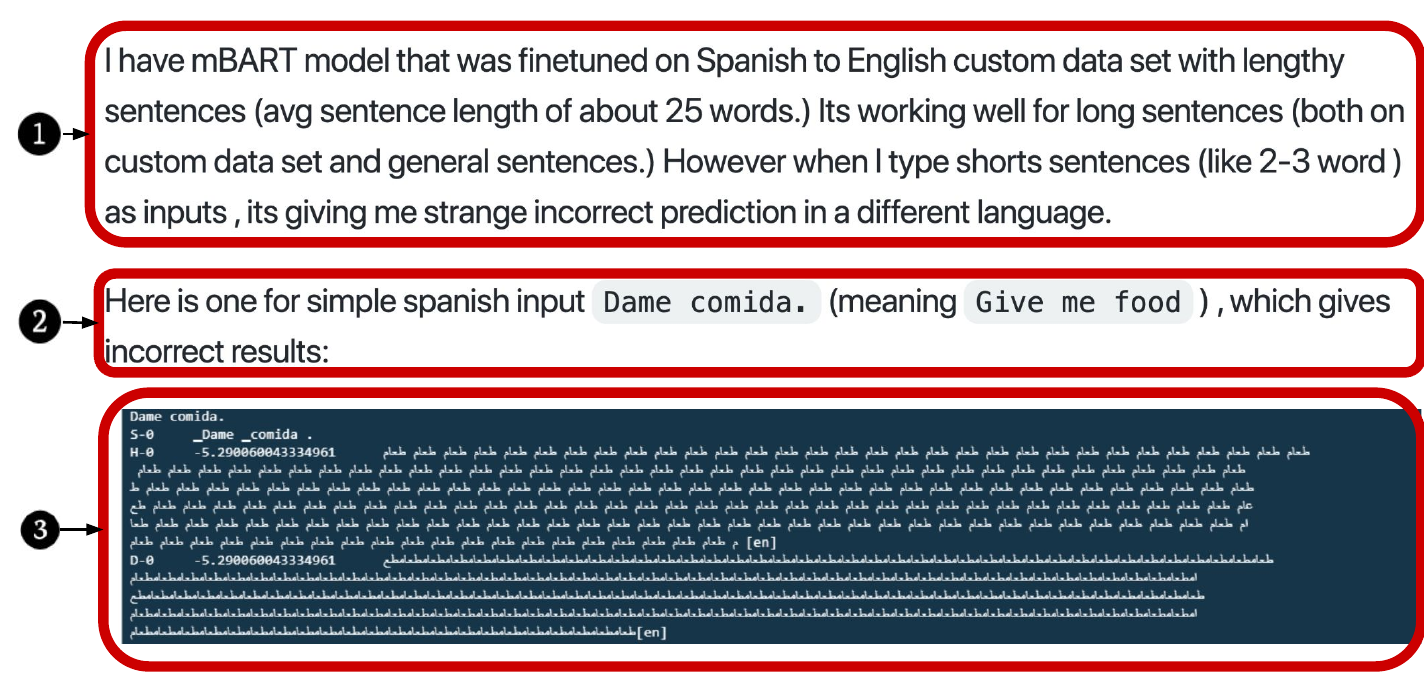}
	\caption{An example of NLP reuse-related bug ~\cite{processing}}
	\label{fig:rq1process}
\end{figure}
The
pre-trained networks analyze the inputs and generates results. Over the last decade, there has been extensive work on processing
natural language and an increasing uptick in usage of these models in both
industries~\cite{devlin2019bert, radford2019language, lample2019cross} and
academia~\cite{vaswani2017attention, tensor2tensor, wolf2020transformers}.
%
\begin{figure*}
	\centering
	\includegraphics[,trim={12cm 0cm 12cm 0cm},width=.7\linewidth]{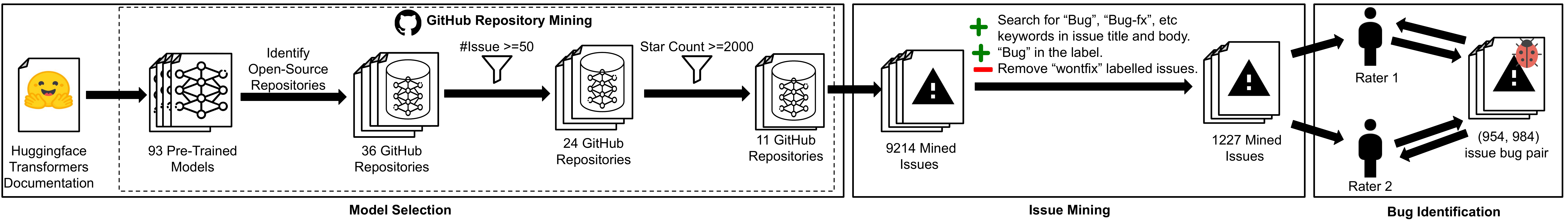}
	\caption{Data collection steps for studying bugs in NLP pre-trained models}
	\label{fig:datacollection}
\end{figure*}
However, NLP models require more training
data~\cite{bender2021dangers} and extensive
computation~\cite{strubell2019energy}, compared to other deep learning (DL)-based
models. The additional computation cost requirements led to the development
of several pre-trained NLP models~\cite{devlin2019bert, radford2019language,lample2019cross} that can be reused with minimal changes.
The extensive reuse of NLP pre-trained models helped many developers implement their desired applications efficiently~\cite{lan2019albert, lewis2020bart, keskarCTRL2019, liu2019roberta}, but the black-box nature of the model often leads to software issues and bugs. 
For example, Figure~\ref{fig:rq1process} shows an issue in which a developer was
fine-tuning a multi-lingual \bart~\cite{bart} model to translate sentences from Spanish to English~(\circled{1}). However, for short sentences, the model is producing output in the Arabic language (\circled{2} and~\circled{3}). Since the developers often do not know the internal processes of the pre-trained model, localizing these issues are challenging. 

There is a vast body of works on understanding the bugs, their root causes, and the
challenges faced by DL~\cite{islam2020repairing,islam2019comprehensive,zhang2018empirical} and machine learning~\cite{thung2012empirical, humbatova2020taxonomy} developers. These works facilitate bug detection, localization~\cite{wardat2021deepdiagnosis, wardat21deeplocalize, nikanjam2021automatic}, and suggest repair strategies~\cite{, zhang2021autotrainer} to practitioners.
However, these studies are mostly done to characterize bugs developers face while building models from scratch, not while reusing pre-trained models. We found that the dataset shared by ~\citeauthor{islam2019comprehensive}~\cite{islam2019comprehensive} has only 16.18\% and 33.57\% of the
DL bugs from \sof and \gh, respectively, are related to reusability. Moreover, the reusability in the prior work's dataset is limited to the image-related models, e.g., ResNet, ImageNet, and are not related to NLP.
In contrast to the prior works, we understand how the bugs introduced while reusing the NLP models are different from the traditional DL bugs and characterized them into the type, root cause, and impact to facilitate further studies. 
In this context, we
consider a bug to be caused by reusing the pre-trained models if (1) the bugs
were present in the reused model and propagated to the code while reusing, and/or (2)
bugs found while adapting the pre-trained model by altering code.


In this study, we curated the 11 most popular NLP
pre-trained models from a model hub, \textit{Huggingface
	Transformers}~\cite{transformer}. Then we mined 9,214 issues from the \gh
repositories of these models that use these pre-trained models. 
We further filtered the issues that report bugs. After manual checking, we
identified 954 issues that contained 984 bugs introduced while reusing NLP models.
We adapt the classification scheme from a prior work~\cite{islam2019comprehensive} to characterize deep learning bugs and update it based on the open-coding scheme. We answer the following research questions: 
\begin{itemize}[leftmargin=*]
	\item \textbf{RQ1 (Bug Types):} What kind of bugs are prevalent while reusing NLP models, and how do they differ from DL software?
	\item \textbf{RQ2 (Root Causes):} What are the root causes of the bugs?
	\begin{itemize}
		\item How do the updates in the pre-trained models introduce bugs?
		\item Which NL configuration parameters are causing the models to be bug-prone?
	\end{itemize}
	\item \textbf{RQ3 (Impacts):} How do the bugs impact the NLP software?
\end{itemize}

%
%

We found several frequent bug patterns that include the wrong update of the models, API misuse, architectural issues, or domain-specific confusion. The key findings
of the paper are as follows:
\begin{enumerate}[leftmargin=*]
	\item Reused NLP models suffer more initialization and memory-related bugs
	than the traditional DL models due to the larger model size. We identified that the average number of parameters and dataset size for the largest DL (ResNet1100, VGG16) models are ~80 million and ~100GB, respectively. Whereas, in NLP (CTRL, T5), it is ~7 billion and ~650GB, respectively, which are significantly large compared to the DL models.  
	\item Traditional DL models are barely reused in production and focus more on data. Whereas, in NLP, the focus is on setting correct parameters (32.3\% of the total bugs) as users often do not have access to the training data of the reused model. 
	\item Due to the black-box nature of the reuse, NLP developers often do not clearly understand the underlying architecture of these models.
	We found that such bugs are seen 14.9\% of the time in NLP, but for DL, it has been seen for 6.3\%.
\end{enumerate}

\textbf{Outline:} 
In \secnref{sec:method}, we discuss data collection and labeling methodology. \secnref{sec:rq1}, \secnref{sec:rq2}, and \S\ref{sec:rq3} describe the frequent bug types, the root causes of the bugs, and  the common
impacts, respectively. \secnref{sec:related} describes the related works, \secnref{sec:threat}
explains the threats to validity, and \secnref{sec:conclusion} concludes.

\section{Study Design}
\label{sec:method}
Here, we describe the methodology to collect the dataset and build
the classification scheme. First, we discuss the data collection steps. Then, we discuss our
classification scheme and labeling approach.
\subsection{Data Collection}
\label{subsec:data}

Figure~\ref{fig:datacollection} shows the data collection steps for selecting the pre-trained models, mining issues from corresponding \gh repositories, and identifying bugs through manual investigation. 

\textbf{Model Selection.} First, we identify the popular open-source NLP pre-trained models and issues corresponding to that. We start by identifying the list of available pre-trained models. For that, we refer to \textit{Huggingface	Transformers}~\cite{wolf2020transformers, transformer}, an open-source framework that maintains an NLP model hub~\cite{modelHub}. It contains
pre-trained models for various NLP tasks such as text classification, translation, etc. Huggingface Transformers is supported by the most popular deep learning libraries (i.e., \tensor, \torch). Furthermore, several major organizations, including Google, Facebook, Microsoft, Allen Institute, and others, use this framework. However, the issues reported in the
\textit{Huggingface Transformers} are related to problems faced by developers that correspond to the bugs in their platform rather than bugs encountered while using pre-trained models. To focus our study on the bugs that developers face while reusing the pre-trained model, we study the issues reported in the repositories corresponding to each such model. As of January 2022, \textit{Huggingface Transformers} listed 93
pre-trained models in their documentation~\cite{transformer}. Among these models, we found that 36 pre-trained models are open-source and have a \gh repository. Since we aim to identify the bugs developers face while reusing these models, we focus on the issues logged by developers in the \gh. Some of these repositories have very few issues since they do not maintain or update the repository based on discussions and reported bugs. Hence, we filter out the repositories having $ \le 50$ issues (open + closed issues). Based on these criteria, we identified 24 repositories. Finally, to focus on the quality repositories (popularity among developers) and keep our manual analysis manageable, we focus on the repositories with at least 2000 stars~\cite{borges2016understanding}. Thus, we found repositories that correspond to 11 NLP pre-trained models. The number of issues for these models ranges from 76 to 3214 and a total of 9214. We further filtered them to identify bugs in these models.
The details of each such model are given in Table~\ref{tbl:dataset}. 

\begin{table*}[!htp]
	\centering
	\setlength{\tabcolsep}{2.2pt}
	\footnotesize
	\begin{tabular}{|p{2.4cm}|p{1.1cm}|c|c|c|p{9.8cm}|c|}
		\hline
		\rowcolor{gray!50}
		\multicolumn{1}{|c|}{\textbf{Model Name}}     & \multicolumn{1}{|c|}{\textbf{Owner}} & \textbf{\#Issue} & \textbf{\#Mined}  & \textbf{\#Bugs}  & \multicolumn{1}{|c|}{\textbf{Description}}  & \textbf{Parameters}     \\
		\hline
		\hline
		ALBERT~\cite{albert, lan2019albert} & GR        & \textbf{171}     & \textbf{64}   & \textbf{53}   & A Lite BERT is trained on multiple datasets, with fewer parameters than \bert.& 223M \\
		\hline
		Bart~\cite{bart, lewis2020bart} & FR         & \textbf{3214}    & \textbf{156}  & \textbf{119} & BART is based on a standard transformer architecture used for neural machine translation. It has both an encoder and a decoder. & 406M \\
		\hline
		Bert~\cite{bert, devlin2019bert}     & GR      & \textbf{1091}    & \textbf{299}  & \textbf{216}  & Bidirectional Encoder Representations from Transformers is a transformer architecture-based model. It has been trained over a lot of unlabeled textual data. & 340M\\
		\hline
		CTRL~\cite{ctrl, keskarCTRL2019}  & Salesforce          & \textbf{76}      & \textbf{33}   & \textbf{27} & Conditional Transformer Language Model  is trained on 140GB of text data. &1.6B\\
		\hline
		GPT-2~\cite{gpt2, radford2019language}  & Open AI        & \textbf{232}     & \textbf{89}   & \textbf{70}   &Generative Pre-trained Transformer-2 is trained on millions of webpages.  & 1.5B
		\\
		\hline
		GPTNeo~\cite{gptneo, gao2020pile}  & EleutherAI        & \textbf{138}     & \textbf{78}   & \textbf{44}  & GPTNeo is a transformer based model trained on 825GiB English text corpus.& 2.7B 
		\\
		\hline
		RoBERTa~\cite{roberta, liu2019roberta} & Facebook      & \textbf{3214}    & \textbf{190}  & \textbf{155}  & Robustly Optimized BERT Approach is an improved version of \bert with more data. &355M \\
		\hline
		T5~\cite{t5, 2020t5}   & Google AI          & \textbf{374}     & \textbf{147}  & \textbf{150}   & Text-to-Text Transfer Transformer model is a large neural network model, trained on a mixture of unlabeled text and labeled data from several downstream tasks. & 11B\\
		\hline
		Transformer-XL~\cite{txl, dai2019transformer}& Google CMU & \textbf{128}     & \textbf{35}   & \textbf{19}  & Transformer-XL model, based on Transformer architecture, allows language understanding without disturbing the temporal coherence unlike traditional transformer. & 257M\\
		\hline
		XLM~\cite{xlm, lample2019cross}    & FR       & \textbf{320}     & \textbf{106}  & \textbf{84}  & XLM is an improved version of BERT that achieves better performance in classification and translation-related tasks.& 500M\\
		\hline
		XLNET~\cite{xlnet, yang2019xlnet}  & Google CMU        & \textbf{256}     & \textbf{80}   & \textbf{47}  & XLNET is a generalized version of \txl. It is based on a large bidirectional transformer that uses larger data. It outperformed \bert on 20 language tasks.& 340M \\
		\hline
		\hline
		
		\multicolumn{2}{|l|}{\textbf{Total}} & \textbf{9214}    & \textbf{1277} & \textbf{984}  && \\
		\hline
	\end{tabular}
\vspace{12pt}
\caption{Dataset description -- Pre-trained NLP models. GR: Google Research, FR: Facebook Research.}
\label{tbl:dataset}
\end{table*}

\textbf{Issue Mining.} Once we fixed the models under study, we mine the
issues related to the \textit{bug} and \textit{bug-fix}. To mine such issues, we
utilize the techniques applied by Garcia~\etal~\cite{garcia2020comprehensive}.
In this approach, the issues with these keywords in the title or body are selected:
\textit{fix}, \textit{defect},
\textit{error}, \textit{bug},
\textit{issue}, \textit{mistake}, \textit{incorrect}, \textit{fault}, and
\textit{flaw}. In addition, the issues labeled as a \textit{bug} by the respective repository are also selected. On top of that, we
remove the issues that are labeled as \textit{wontfix}, which are
primarily not bugs, and there is no fix needed for those issues. Finally, we
identified 1227 issues from \gh out of 9214 issues present in all ten
repositories.

\textbf{Bug Identification.} Once we mine all the bugs-related issues
based on the keyword searching, two raters (first and second authors)
manually went through the issues to mark them as \textit{``bug''} or
\textit{``not bug''}. If there is a discrepancy while labeling the issues, it has been clarified based on the discussion among raters. Finally, 954 
bug-related issues are identified from 1227 issues. Within
these 954 issues, 30 bug-related issues have more than one bug in a single issue.
For such scenarios, we counted them as more than one bug. 

\subsection{Classification Strategy}
\label{subsec:classification}
Here, we classify the bugs into their types (bug-type), what is causing the bug (root cause), and how the bug affects the system (impact). We define the classification scheme based on prior works~\cite{islam2019comprehensive, zhang2018empirical, beizer1990software} and follow the open coding scheme for each category. 
First, a pilot study has been conducted to identify the need for a new category, and if needed, it has been approved in the discussion among the raters. For impacts and bug types, we found that the classification scheme proposed by \citeauthor{islam2019comprehensive}~\cite{islam2019comprehensive} and \citeauthor{zhang2018empirical}~\cite{zhang2018empirical} are sufficient to address bugs found in NLP-based systems. 
For root cause, we adapted the classification scheme proposed by \citeauthor{islam2019comprehensive} and added new root cause kinds based on the open coding approach. We have added five main kinds and several sub-kinds within that. The entire classification scheme has been shown in \fignref{taxo}.

\textbf{Labeling}. We label the bug-related issues based on the classification scheme fixed in the previous step. The first and second authors independently labeled the bugs. To measure the raters' agreement, we compute the Cohen's Kappa co-efficient~\cite{viera2005understanding} for each 10\% of issues. After two rounds, the Cohen's kappa coefficients for all three categories were $>0.85$, which indicates the perfect agreement. The first and second authors then labeled independently and resolved any disagreement based on discussions.  


%
\section{What are the prevalent bug types while reusing NLP models}
\label{sec:rq1}

Here, we discuss the type of bugs found while reusing NLP models.

\textbf{API Bug.} 
API-related issues are mostly due to the
changes in the top-level APIs that cause the models to behave aberrantly. 
These types of bugs can occur both from misuse by the developer of the client code as well as for missing the compatibility among the
APIs. 
%

\textbf{Data Bug.} When bugs are caused while loading the training data for the NLP models, we refer to them as data bugs. 

\textbf{Structural Bug (SB).} This genre of bugs is associated with the
incorrect definition of the model structure. 
Bugs associated with the structure of the model is classified further into five categories:
\begin{enumerate}[leftmargin=*]
	\item \textbf{Control and Sequence Bug:} These bugs are related to the control sequence or control flow of the code.
	\item \textbf{Data Flow Bug:} While data bugs are related to the input data,
	data flow bugs are referred to as the bugs seen when data passes through the
	model structure. For instance, the output shape of the data after applying a layer operation does not match with the input required for the operation of the consecutive layer. In such cases, we classify the bugs as data-flow bugs.
	\item \textbf{Initialization Bug:} Reusing the NLP models involves the initialization of
	different hyper-parameters. These parameters help define the vocabulary size, the sequence length of the input words, the number of sentences in a single input to the system (batch size), etc. Often wrong initialization can cause the NLP model to crash, produce unexpected outcomes, etc.
	\item \textbf{Logic Bug:} Bugs can occur due to logical errors in the model and the code
	structure. 
	\item \textbf{Processing Bug:} These bugs are related to the wrong choice of the algorithm. 
\end{enumerate}
\textbf{Non-Model Structural Bug (NMSB).} Any bugs (except data and API bugs)
initiated outside the model are classified as ``non-structural bugs''. The sub-categories are the same as structural bugs. However, we only find initialization-related bugs in this category.
\subsection{What are the key differences between DL and NLP bug types?}

We found that the bug-type categories proposed by \cite{islam2019comprehensive} are sufficient to represent the bugs seen while
reusing the NLP pre-trained models. However, the distribution, characteristics,
and implications are significantly different. 
Here, we discuss each such category and extend that to highlight the
commonalities and variabilities with the DL bugs. 
\begin{figure}
	\centering
	\includegraphics[width=.95\linewidth]{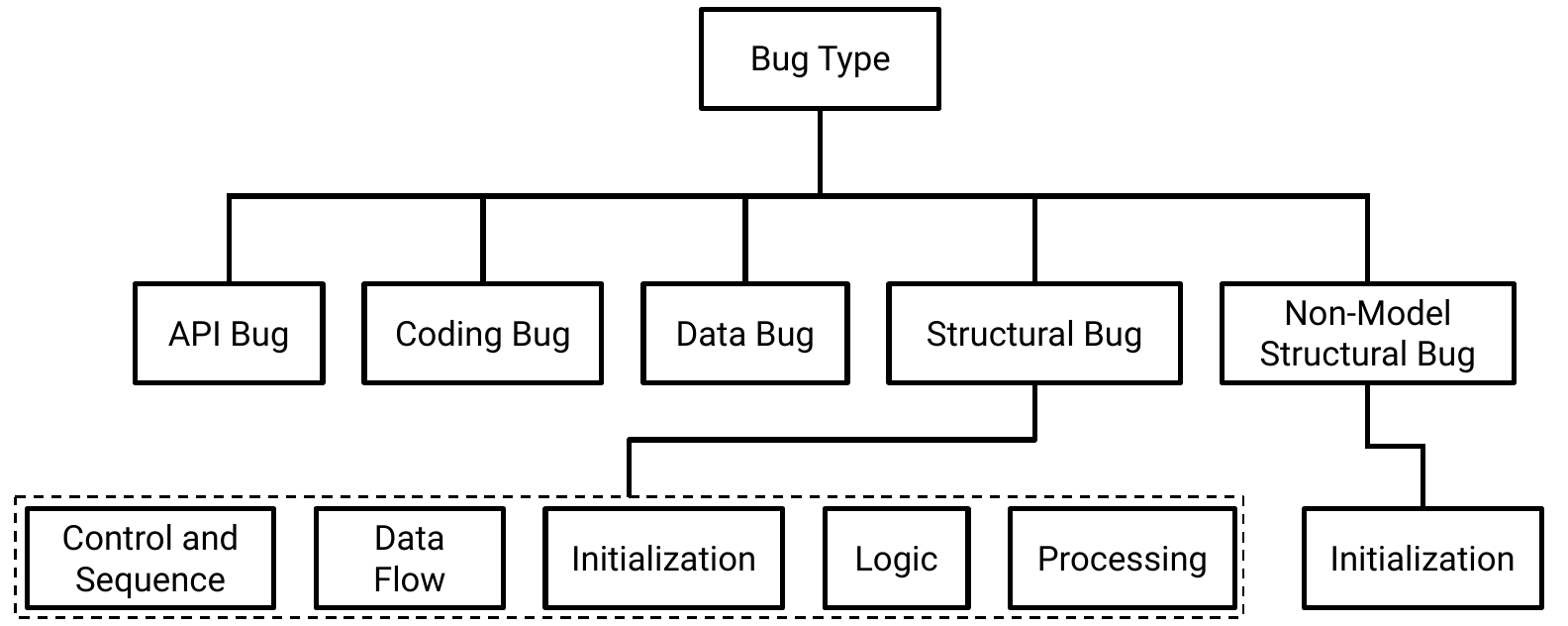}
	\caption{Classification of bug types}
	\label{fig:bugtypeTaxo}
\end{figure}
\begin{figure*}[!htp]
	\centering
	\begin{subfigure}{.5\textwidth}
		\centering
		\includegraphics[, trim={2cm 7cm 0cm 5cm},width=0.55\linewidth]{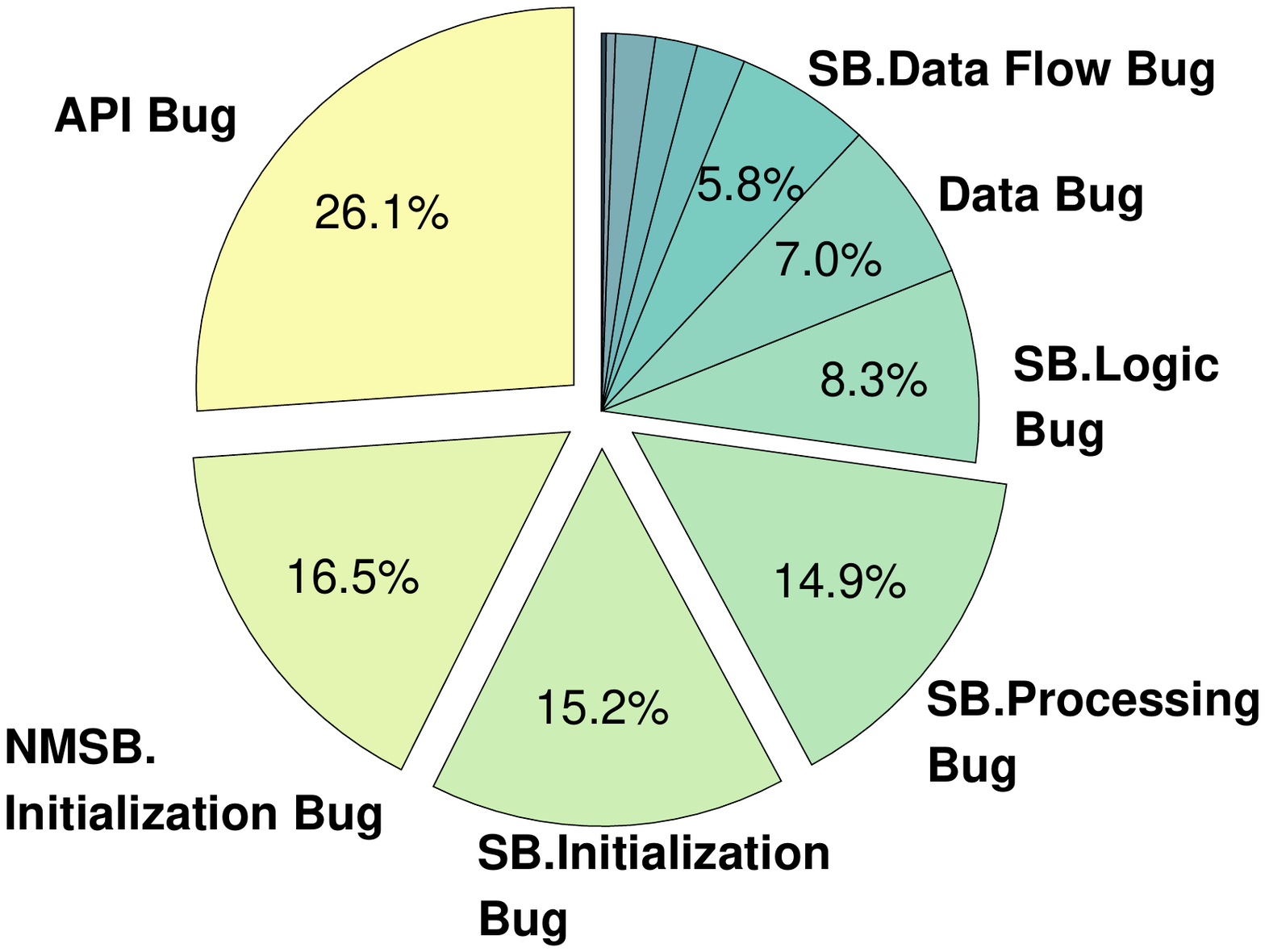}
		\centering
		\caption{Distribution of bug types in NLP (labels $<$5.8\% are hidden)}
	\label{fig:bugtype}
	\end{subfigure}%
	\begin{subfigure}{.5\textwidth}
		\centering
		\includegraphics[, trim={2cm 7cm 0cm 5cm},width=0.55\linewidth]{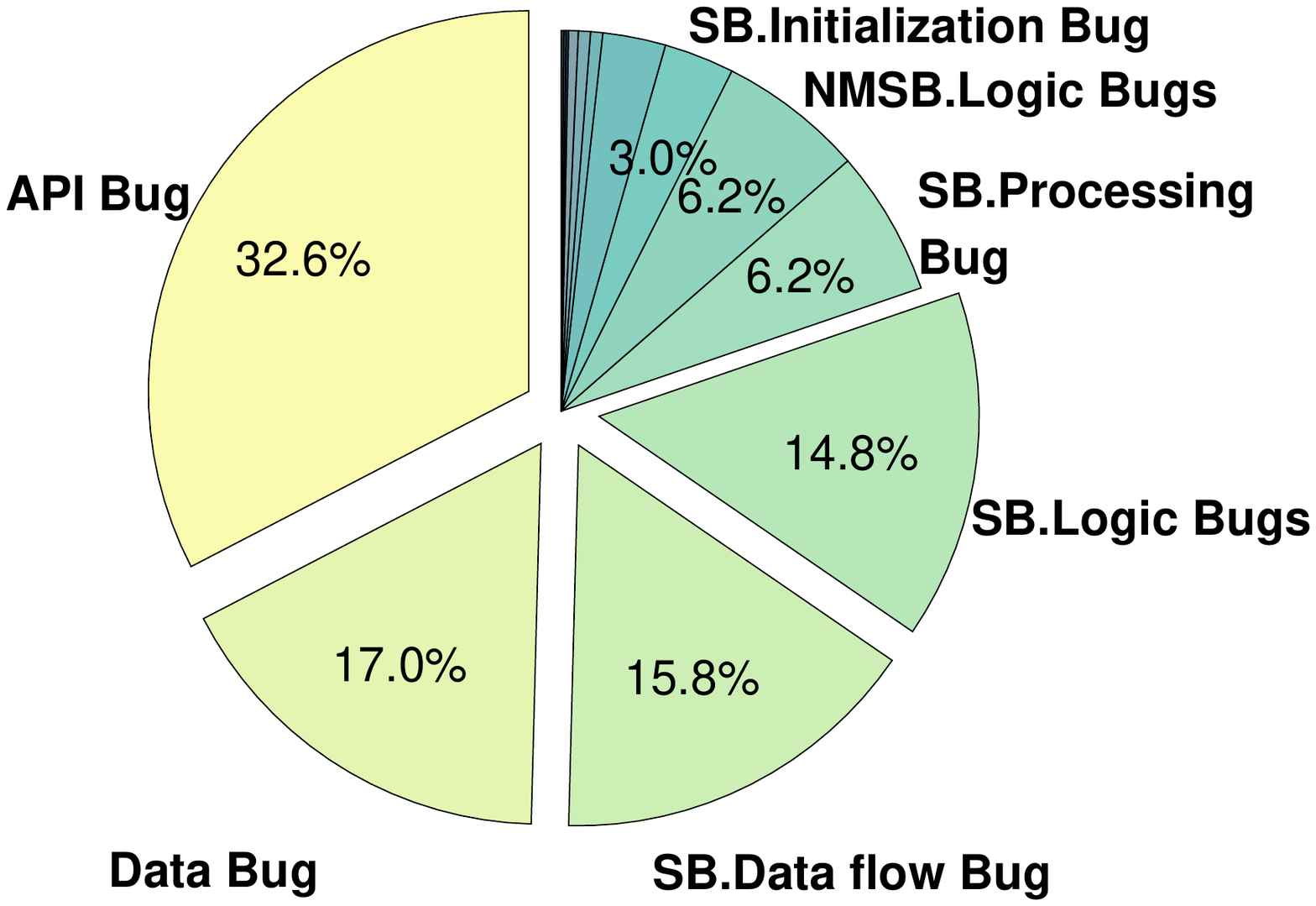}
		\caption{Distribution of bug types in DL (labels $<$3.0\% are	hidden)}
	\label{fig:bugtypeDL}
	\end{subfigure}
	\caption{Comparison between NLP and DL bug types~\cite{islam2019comprehensive} and their distributions}
\end{figure*} 
In Figure.~\ref{fig:bugtype} and \ref{fig:bugtypeDL}, we show the distribution of bugs in NLP-based software and the traditional DL-based software, respectively.
We found that API bugs (26.1\%), initialization bugs (both structural and
non-structural) (31.7\%), and processing bugs (14.9\%) have been observed for more than 70\%
of the bugs in our dataset. Whereas, in the traditional DL, the most prevalent bugs are API bugs (32.6\%), data bugs (17.0\%),
dataflow bugs (15.8\%), etc. 

This suggests that both NLP and traditional DL software are API intensive. We found that versioning (16.02\%), API incompatibility (23.44\%), and execution environment setup (15.63\%) are the most common cause of API bugs. 
A uniform API definition can help inform the developer about different model reuse. For example, even though Huggingface provides a platform to reuse curated models, they do not provide a unified API specification applicable to different models. Building such unified specifications could help developers to avoid these issues.

We also found that the
dependency on the data for NLP-based software and DL-based software is significantly different. First, a pre-trained NLP
model requires little to no additional data for further usage. Second,
retraining these NLP models is not as data-intensive as training a traditional DL
model from scratch. Here, the majority of
the obstacles developers face are setting the correct model for a problem
(NLP: 14.9\%, DL: 6.2\%) and correctly reusing the model (NLP: 31.7\%, DL: 3.5\%), and such issues rarely occur in the traditional DL-based software. 
Also, the logic bugs (8.3\%) are not as common as in traditional DL software (18.1\%). The primary reason is that developers often do
not alter pre-trained models' logic. 
Next, we discuss some of the key differences.


\subsection{How Initialization Bugs in NLP are Different From Traditional DL Bugs?}
One key part of building a model in both DL and NLP-based software is the choice of the parameters, setting up the execution environment, etc. We found that the majority of such problems are due to the massive size of these pre-trained models.
\finding{Larger NLP models introduce more initialization bugs.}
Initialization bugs are prevalent in all models, with a heavy presence in \gpt and \ctrl. 
Further investigating, we found that 25.47\% of the bugs in this category originated due to issues while setting up the execution environment. In the traditional DL model, initialization bugs do not occur much (3.29\%).  Since the average size of the DL model is significantly less than that of NLP models, the wrong environmental setup does not affect as much as it does in NLP. We found that for the largest DL models (ResNet1100, VGG16), there are $\sim$80 million parameters, and the size is $\sim$100GB, on average. Whereas, for the same NLP models (CTRL, T5), the average number of parameters and size is $\sim$7 billion and $\sim$650GB, respectively, which is significantly more than the DL models. 
In Table~\ref{tbl:dataset}, the models with their corresponding sizes are shown. To identify the model size, we looked into various versions of the models and listed them according to the total number of parameters. 
Due to the large size of the model, loading them to CPU or GPU requires high resources. These bugs cause the abrupt stop of the
execution (66.67\%) and memory issues (13.65\%). Also, Strubell~\etal~\cite{strubell2019energy} identified the resource consumption of the models and their impact on the environment. They found that training a language model emits 6x more $CO_2e$ than a car running on fuel for a year. So, while using the models, the size
should also be taken into account. 

\textbf{\textit{Implications.}}
The reuse of pre-trained NLP models is very similar to the software reuse before the notion of modular programming. With finer reuse~\cite{parnas1976design, parnas1972criteria}, parts of the software can be reused and replaced. Recently, for image-based classification problems, \citeauthor{pan2020decomposing}~\cite{pan2020decomposing,pan22decomposing} have proposed an approach to decompose a monolithic model into modules to enable reusability and replaceability of the decomposed modules. Similarly, a more modular architecture can be proposed for the NLP-based systems so that instead of reusing the entire model, developers can use the parts of it.
\subsection{How processing bugs in NLP are different from traditional DL bugs?}
Compared to the DL, NLP-based software suffers from bugs related to the correct algorithm and parameters to retrain the model.
\finding{14.9\% of the bugs in reusing models are related to the
	processing bugs.}
For instance, previously, we discussed an issue
(Figure~\ref{fig:rq1process}), where the developers could not finetune
the model for the language translation task (Spanish to English). The
translation works perfectly if the sentence length is large.
However, when the sentences are short, it predicts in different languages other
than the languages specified while finetuning the model. While the model might have
the knowledge of the third language, that is unnecessary for
the developer's need. Without knowing the underlying architecture, fixing such bugs
are not always possible. In fact, the situation is akin to what has been proposed by \citeauthor{parnas1976design}~\cite{parnas1976design} regarding the program family. In that work, Parnas argued that software could be reused either as a complete
program (that produces output given input) or as smaller
components (intermediate stages that may or may not produce output given input).
While reusing the complete program, the traits of
the ancestors are passed to the descendent program without the knowledge. Some of these traits are important, but not all. In the
previous example, the same happened. While reusing, the
descendent software receives knowledge of another language other than English
and Spanish, which is not necessary for this context. Such bugs are not that common in traditional DL, as reusability is not as common as in NLP-based software.
For traditional DL models, such bugs only account for 6.3\% of the total bugs. While investigating, we found that such bugs in traditional DL are due to confusion while choosing the correct algorithm to train the model. However, for NLP, it is mostly the black-box nature of the reuse.

\textbf{\textit{Implications.}} 
To avoid such scenarios,
Parnas~\cite{parnas1976design} has suggested that one needs to reuse the intermediate stages of the
software and reuse only the required
information. While the notion of the complete program vs. intermediate program
has not been identified yet, work by~\citeauthor{pan2020decomposing}~\cite{pan2020decomposing, pan22decomposing} has
identified how these black-box models can be seen as a composition of
smaller black-boxes. Understanding these smaller black-boxes might help
researchers capture the model's intermediate representation and reuse that
instead of the complete models.
%
%

\section{Common Root Causes}
\label{sec:rq2}
	\begin{figure*}[]
	\centering
	\includegraphics[width=1.0\linewidth]{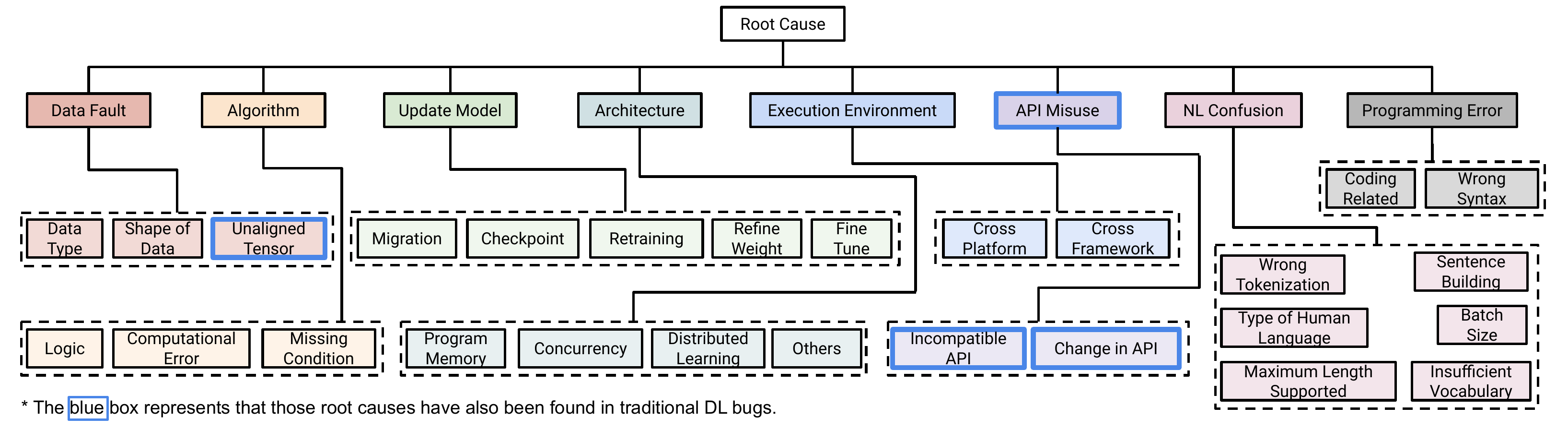}
	\caption{Classification of root causes of bugs in NLP pre-trained models}
	\label{taxo}
\end{figure*}
In this section, we discuss the common root causes of the bugs. First, we discuss all the root causes that we identified, then highlight the most prevalent causes. In Figure~\ref{taxo}, we illustrate all classifications of the root causes, and the blue boxes represent that these bugs are also present in the traditional DL models.

\textbf{Data Faults.}
If the bugs are caused by the incompatibility of the data to the model, we categorize them as data faults.
%
Bugs in this category can be broadly classified into, 
\begin{enumerate}[leftmargin=*]
	\item \textbf{Type:} These bugs occur due to wrong or mismatched data type.
	\item \textbf{Shape:} Dimension mismatch of input data leads to
	these bugs.
	\item \textbf{Size:} Often, the model expects a predetermined size of the input,
	which can cause these bugs.
	\item \textbf{Unaligned Tensor:} Incompatibility in the data flow of the
	tensors can cause bugs.
\end{enumerate}

\textbf{Algorithmic Error.}
Bugs can occur due to logical or conditional errors such as missing conditions in weight updates, division by zero, etc. The root causes are sub-categorized into,
\begin{enumerate}[leftmargin=*]
	\item \textbf{Logic:} These bugs occur due to
	missing any concept or logic. 
	\item \textbf{Computational Error:} Such bugs appear when existing
	computation produces incorrect results. 
	\item \textbf{Missing Conditions:} If the pre- or post-conditions of any computation
	need to be added/fixed.
\end{enumerate}


\textbf{Updating the Pre-trained Model.}
Reusing the pre-trained models often requires satisfying model-specific
requirements. Violating them can introduce bugs in the system. For example, 
\begin{enumerate}[leftmargin=*]
	\item \textbf{Migration:} While migrating pre-trained models to accommodate different
	tasks or datasets, the model requirement does not match with other environments and causes bugs. 
	\item \textbf{Checkpoint:} In DL, creating checkpoint is a common practice during training. This helps one to get intermediate results while performing extensive training. However, we found that often problems may occur while saving these checkpoints or accessing existing checkpoints of the trained model.
	\item \textbf{Re-train :} Bugs introduced while re-training the models on new training data.
	\item \textbf{Refine Weights:} A pre-trained model can offer
	ways to
	refine existing weights. 
	\item \textbf{Fine Tune: }Bug caused while tuning the parameters.
\end{enumerate}

\textbf{Architectural Incompatibility.}
The prebuilt NLP models can have an implicit architectural requirement that poses different
types of incompatibility in the user's system. 
Moreover, the models are resource
hungry and sometimes do not go through exhaustive testing on different
combinations of computational architecture, which causes these bugs. 
The types of such bugs are: 
\begin{enumerate}[leftmargin=*]
	\item \textbf{Memory:} The
	pre-trained models require high memory and often that causes bugs.
	\item \textbf{Concurrency:} Many models allow
	multiple training threads for speedup but cause bugs for strict concurrency assumptions. 
	\item \textbf{Distributed Learning:} Sometimes, these models can be run on the distributed architecture, e.g., multiple cores of CPU or GPU, which leads to bugs.
\end{enumerate}

\textbf{Incorrect Execution Environment.}
Pre-trained models have been made available within different DL packages, e.g.,
\tensor, \torch. 
The subtypes in this category are:
\begin{enumerate}[leftmargin=*]
	\item \textbf{Cross-platform:} These bugs can occur because of specific requirements of
	the underlying platform, e.g., operating system. 
	\item \textbf{Cross-framework:}
	Incompatible versions of Python or required packages such as \tensor, \torch can be the root cause of this type of bug.
\end{enumerate}

\textbf{API Misuse.}
These bugs can be caused due to, 
\begin{enumerate}[leftmargin=*]
	\item \textbf{Absence of Inter-API Compatibility}:
	There are many API dependencies in the NLP programs, i.e.,
	\textit{Scikit-Learn}, \textit{Tensorflow}. Often these APIs are not compatible in performing a task jointly.
	\item \textbf{API Change:}
	These bugs are related to API versioning.
\end{enumerate}


\textbf{Confusion in NL-specific Specifications.}
A vast majority of the bugs are due to the missing vocabulary, incorrect sequence length, wrong choice of tokenization mechanism while
preprocessing the data and re-training the model. We further categorized such root causes into the
following categories:
\begin{enumerate}[leftmargin=*]
	\item \textbf{Vocabulary: }Bugs can occur because of
	inappropriate vocabulary or accessing the vocabulary in specific tasks.
	\item \textbf{Tokenization: }When re-training the models on a new dataset, the
	tokens should be preprocessed in accordance with the pre-trained model versions.
	\item \textbf{Sentence: }Often, the semantics and syntax of the sentences differ
	with the customization of NLP tasks, which leads to bugs. 
	\item \textbf{Batch Size: }Because of heavy resource computations and lengthy
	training process, the batch size needs frequent updating, which results in
	NL-specific bugs.
	\item \textbf{Language:} Since different human languages have a variety of structures;
	specific preprocessing and task customization is needed to reuse the pre-trained
	models. In this process, developers can face NL-related bugs.
	\item \textbf{Maximum
		Length: }These models have a hard threshold of the supported maximum length of
	sentences. Developers often need to tune that, which causes bugs.
\end{enumerate}

\textbf{Programming Error.}
These bugs occur due to any other programming issues such as: 
\begin{enumerate}[leftmargin=*]
	\item \textbf{Syntax
		Error:} Bugs might occur due to incorrect syntactical errors such as missing
	punctuations,
	parenthesis, etc.
	\item \textbf{Coding:} Other programming errors such as the wrong loop
	breaking, missing corner cases, misusing variable names, etc. causes bugs in
	the NLP models.
\end{enumerate}
\begin{figure*}[!htp]
	\begin{subfigure}{.5\textwidth}
		\centering
		\includegraphics[, trim={2cm 7cm 0cm 5cm},width=0.5\linewidth]{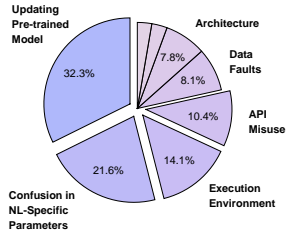}
		\caption{Distribution of root causes in NLP (labels $<$7.8\% are
			hidden)}
		\label{fig:rootcause}
	\end{subfigure}%
	\begin{subfigure}{.5\textwidth}
		\centering
		\includegraphics[, trim={2cm 7cm 0cm 5cm},width=0.5\linewidth]{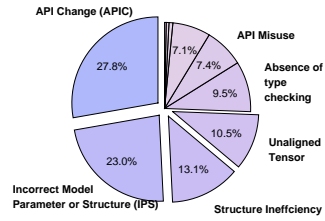}
		\caption{Distribution of root cause types in DL (labels $<$7.1\% are hidden)}
		\label{fig:rootcauseDLs}
	\end{subfigure}
	\caption{Comparison between the root causes of NLP and DL bugs~\cite{islam2019comprehensive} and their distributions}
	\centering
\end{figure*} 
We found that the root causes are significantly different and very specific to the NLP-based software.
In Figure~\ref{fig:rootcause}, we show the high-level distribution of the different root causes. We also show the distribution of root causes that are prevalent in traditional DL-based software in Figure~\ref{fig:rootcauseDLs} based on bugs reported by \cite{islam2019comprehensive}. Comparing both figures, we can observe that the types and the distributions of root causes are significantly different. For instance, DL models are more prone to bugs when API changes. However, since NLP models are pre-trained with a version of API, users do not tend to change the version, resulting in fewer bugs. Also, most bugs in the NLP system are updating the model by retraining, refining weights, etc., which are not common in DL-based software. This section focuses on the main root causes of bugs in NLP-based models, which are not present in the traditional DL. 
%
\subsection{How updating the pre-trained models introduce bugs?}
\label{subsec:reuse}
Approximately one-third of the bugs in reusing pre-trained models are related to updating models. Such circumstances are very nominal in traditional DL, with prior work not defining any classification category representing their root causes. In this context, updating a pre-trained model can be done by migrating the model to another system, setting up checkpoints to store intermediate training results, retraining the model with either additional or new data, refining the weight, fine-tuning the model, etc., to match the user's intended functionality with the one provided by the model.

\finding{Majority of the bugs (32.3\%) are caused due to updating the pre-trained models.}

In an issue reported by a developer~\cite{rq2f1} discusses that re-training a \tf model with a custom dataset and the model produces unexpected
output, generating prefixes in German or Spanish while the model was
trained for the English dataset. This is due to insufficient training. Further investigating, we found
several sub-categories of such bugs, e.g., migrating models and model-related artifacts,
fine-tuning, imposing checkpoints, refining models' weight and bias, and
re-training models with custom datasets and parameters. 
As discussed previously, this phenomenon is unique to the NLP model reusability
as it is not very common to reuse traditional DL models. Most traditional DL models are built from scratch. For instance, only 3.5\% of the models curated (out of 6368 \gh repositories) by \citeauthor{gonzalez2020state}~\cite{gonzalez2020state} are related to reusing pre-trained model. We verify the reuse by identifying the presence of a pre-trained model in the list of import statements, and/or a compiled model is loaded to the
code. We found that most bugs introduced while re-training are due to migrating models, re-training the model, and fine-tuning the models, and we discuss each cause in the paragraphs below.

\subsubsection{Migration}
\finding{11.17\% of the bugs are introduced while migrating the pre-trained
	models.}
A majority of the bugs are generated while loading
the pre-trained models and other associated components. The most common
ones are the missing files and the incompatibility between the model and the
input. For instance, an issue~\cite{migration} in
\bert, where a conversion error occurred due to the mismatch between the input
requirement of the pre-trained model and the available APIs, e.g., \tensor.

\textbf{\textit{Implications.}}
These migration issues are very similar to the software component migration~\cite{plakidas2018software, kum2008autosar, fleurey2007model, choi2004fast,phan2017statistical} problems, where due to ever-changing platforms, paradigms, and techniques, transferring software from one infrastructure to another may cause bugs. However, approaches to resolving such issues~\cite{plakidas2018software, kum2008autosar, fleurey2007model} are limited to the traditional software and currently are not applicable to both DL and NLP-based software. Also, in SE, there have been works on program fragments and linking~\cite{cardelli1997program}. In this particular work~\cite{cardelli1997program}, Cardelli has proposed the notion of the separate compilation of the program fragments and linking them together to form a complete program. The components involved in the migration tasks, e.g., the model, input data, API, can also be compared as program fragments, where each fragment cannot individually be compiled or type-checked. The type-checking only occurs when all the components form the complete program. If the type-checking fails while linking the fragments, then migration-related bugs occur. However, suppose we can enable the separate compilation of the program fragments,  we can identify whether each fragment can safely be linked to other components or fragments. Also, we can validate if certain fragments can be safely replaced with other fragments while migrating to a different environment.  We believe that both research directions could be an interesting avenue to venture into the SE-PL-DL community.

\subsubsection{Retraining} 
\finding{9.14\% of the all the bugs are related to retraining.} 
The pre-trained models can be reused (1) without any modification, (2) by re-training
with changing parameters, or (3) by re-training with an additional dataset. In this
section, we discuss the second and third approaches of reusing pre-trained
models.
Re-training without new data can be achieved when a model $M$ is trained on a dataset $D$ and has been further trained with a different set of initialization
parameters. Prior research~\cite{song2020information, zanella2020analyzing} has found that retraining increases the chance of information leakage as the dataset is not evolved during the process. So, if an adversary receives certain information about the dataset, the NLP model can easily be attacked by perturbing the words~\cite{zhang2021red}.

Retraining by addition of dataset is also referred to as incremental
training~\cite{syed1999incremental} \cite{wu2019large} or continuous
learning~\cite{collobert2008unified}. Here, a model $M$ is trained with a dataset
$D$, and re-trained with new data $D'$ to create a model $M'$. For instance,
in an issue~\cite{retraining} logged by developers discussed that while using \bert
model on an example dataset, the accuracy is very low. However, while
predicting, the prediction accuracy is very high (close to 1) for the testing
dataset. To fix the issue, the model requires more training. Since the pre-trained model already has a
certain amount of knowledge, it was sufficient to predict the
examples in the testing dataset, which is essentially smaller than the training dataset. However, it is not sufficient for predicting
examples from the training dataset.

\textbf{\textit{Implications.}}
Developers are often not aware of the differences between different types of
re-training. The focus is more on accuracy than on other aspects, e.g.,
fairness and data leakage. Updating the APIs with the consequences of different
re-training can help developers make better-informed decisions.

\subsubsection{Fine Tuning}
Developers often re-structure the pre-trained models to fit the requirement. For
instance, in an issue reported in \bert~\cite{finetuning}, where the developer modified the model, which mostly includes steps, i.e., pruning the model
changing the parameters. However, it has been noticed that fine-tuning does a negative effect due to the presence
of spelling mistakes. A single
mistake from ``could'' to ``cud'' has changed the prediction result as well as the
attention on the words. Such incidents could also be used as backdoor attacks to the model, where an attacker can knowingly change a single word in such a way that the final outcome of the model has also been changed~\cite{chen2021badnl, yang-etal-2021-rethinking}.

\subsection{What are the most bug prone NL-specific parameters?}
\begin{figure}[!htp]
	\centering
	\includegraphics[width=0.8\columnwidth]{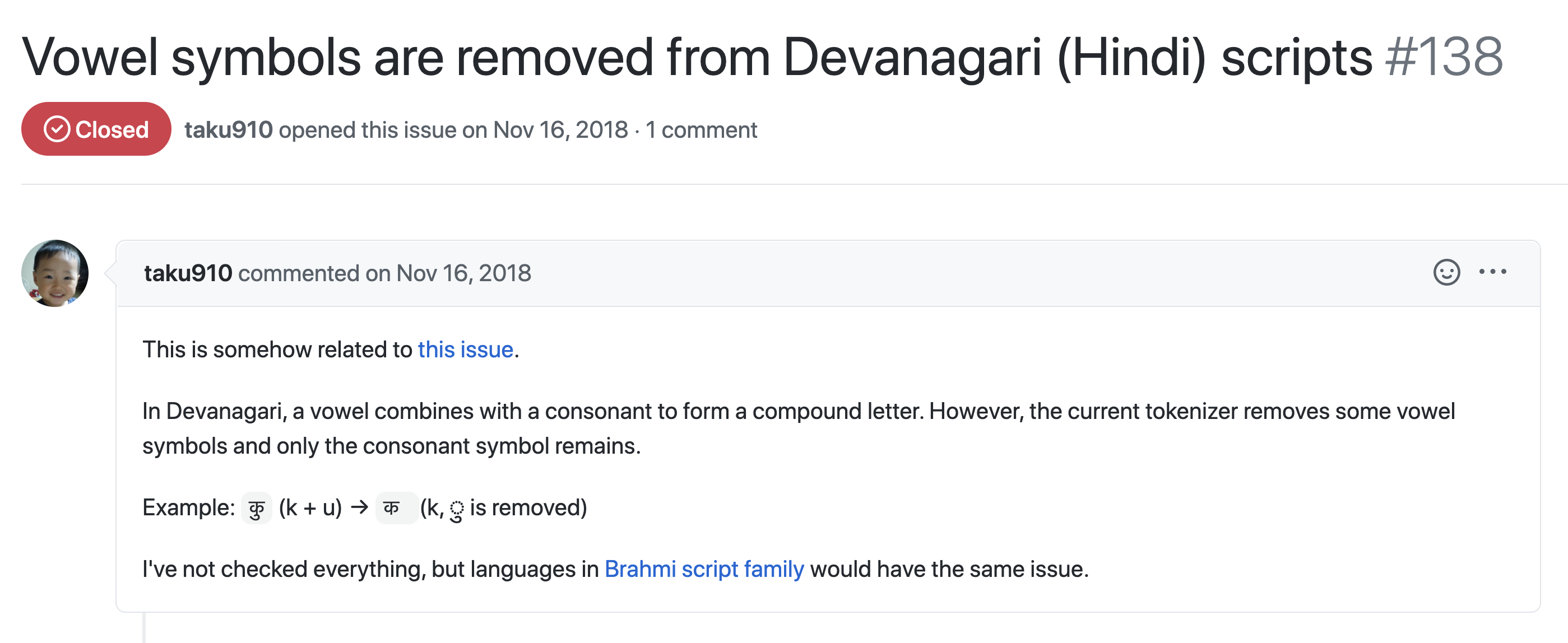}
	\caption{Example of the wrong specification~\cite{cnls}}
	\label{fig:cnls}
\end{figure}
A majority of the bugs (21.4\%) are caused by using a wrong parameter
while reusing the NL models. Most such cases are caused due to the black-box nature of these models. Developers choose to tune these parameters without knowing the internals of the model. Moreover, such practices often go beyond the accuracy of the model and impact on robustness, fairness, and other properties of the model. Here, we discuss these NL-specific parameters and other specifications, e.g., tokenization, language, vocabulary; developers can alter that. Also, we describe how different choices would impact the model's behavior. For instance, Figure~\ref{fig:cnls} shows an example where the developer is having trouble reusing the \bert model to perform a certain task in the Devanagari language. There is a concept of compound letters in this particular language, where two vowels can be combined to form another letter (as suggested by the developer in the issue). Since such knowledge is not present in the reused model, the tokenizer only works on the first vowel of the compound letter. Such bugs are related to the wrong specification of the upstream and the downstream software. Here, we discuss such types of root causes in detail. 

\finding{Wrong NL batch size and sequence length can affect the robustness of the NL software.}

\subsubsection{Batch Size}
Among all the NL-related parameter settings, setting up the correct \textit{Batch Size} is the cause for most of
the bugs (21.76\% of bugs in this category (4.65\% overall)). Batch size controls the flow of the input to the model while training. Every model is shipped with a default value for this parameter. However, to accommodate the model and execute on the tailored problems, one might need to alter the value of this parameter. The value can either be increased or decreased based on the need and the available resources. However, both options have their pros and cons.

\textbf{Decreasing the Batch Size.} 
36.96\% of the bugs caused by the batch size end up having memory
issues, and the other 34.78\% halts the program abruptly. We found that all the bugs
related to the memory issue have a common fix: decreasing the batch size to
accommodate the program with limited resources. However, developers are unaware that, while decreasing the batch size will reduce the memory consumption
and make the learning process faster, it might also decrease the robustness. If the NLP
model is related to a safety-critical system or dataset with sensitive
information, then decreasing the batch size will jeopardize the system's safety. NLP, as well as
most DL architectures, use a stochastic gradient-based approach for
learning. With a fixed batch size, the gradient computation approximates the
loss between two input batches. If a developer decreases the batch size
significantly, then computed approximation will have high variance, and it will
direct the model to learn faster. Small batch size often helps skip some local minima
and contribute to the right direction~\cite{mccandlish2018empirical}. However, if the batch size is
extremely small, then two things can happen. First, if the dataset has a bias
towards a particular word, then sample bias could be created, and the model will
try to remember the word. If an adversary changes the word to a different word,
then it will decrease the overall robustness of the
model~\cite{galloway2019batch}. For instance, there is a sample bias over the
word ``dog'' in a dataset. 
If a sentence, ``Dog is barking'' is classified as negative and the
adversary only changes the ``barking'' with ``eating'',
then the sentence will be classified as negative as well, which is incorrect. Here, due to the small batch size, the model starts to remember the words associated with the sentiment and predicts if there is a presence of such words without validating other sections of the sentence. Second, if the batch size is very small, the model might never converge and will never halt, and thus, it will not
reach the desired accuracy, which will make the NLP system vulnerable to adversarial attacks.

\textbf{Increasing the Batch Size.} 
While decreasing batch size will
decrease the robustness, increasing the batch size can also have adverse
effects. If the batch size is too big, then the approximation
between the two batches of input is too large, which will not help
the model learn and take more iteration to converge. This is the
primary reason for having memory out of bound error. 

\textbf{\textit{Implications.}} There is no optimum value for the batch size that can help the
model be more robust. Runtime verification could be done on the gradient
approximation to identify such issues. This could be done at either API level,
e.g., TensorFlow and PyTorch APIs, or by building a
dynamically analyzing gradient approximation. If the value
exceeds a certain threshold, developers could either be informed or the program
can be halted.

\subsubsection{Sequence Length.}
\begin{figure}
	\includegraphics[width=1\linewidth]{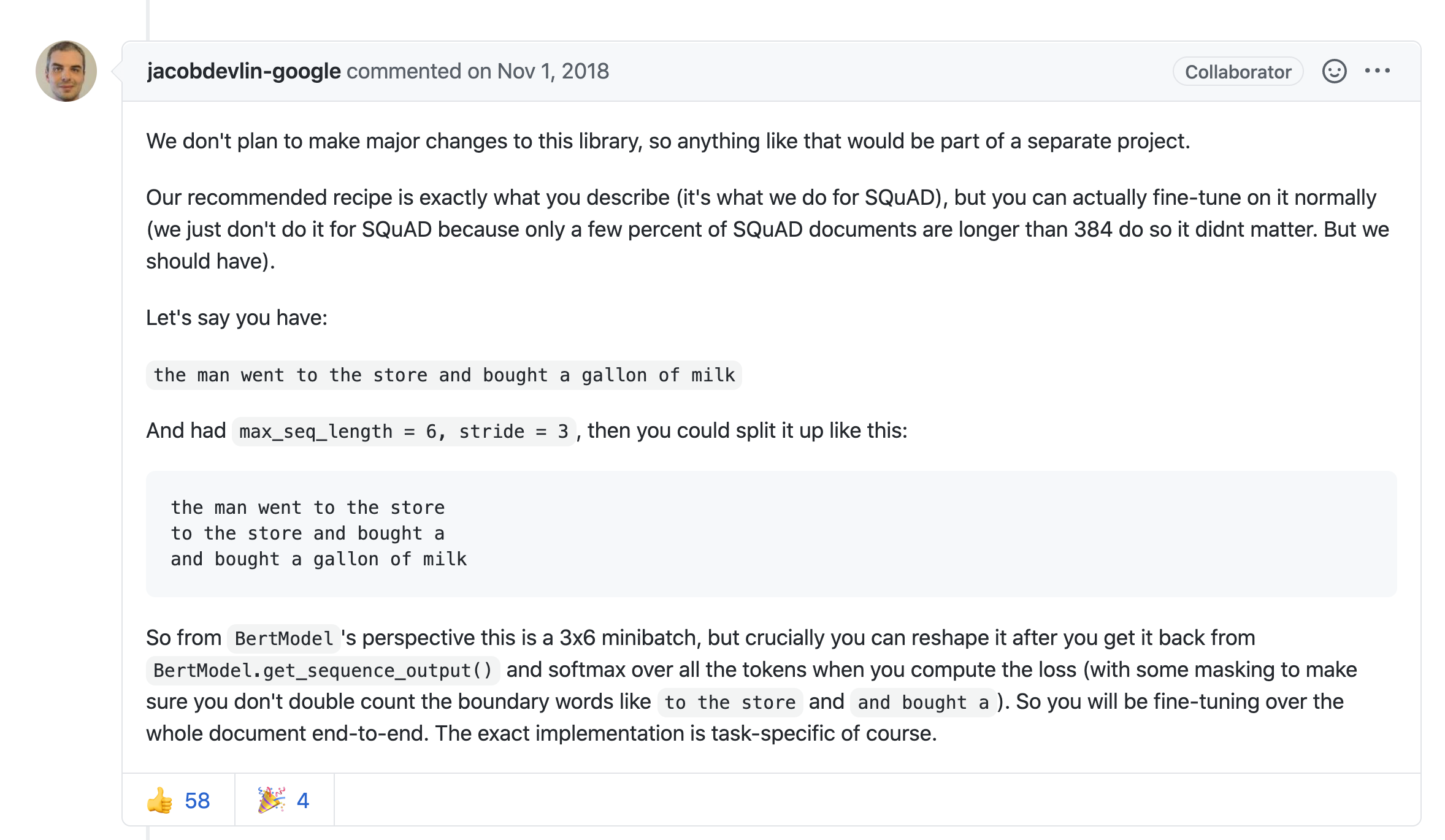}
	\caption{Solution to a max length related bug~\cite{max}}
	\label{fig:max}
\end{figure}
16.20\% (3.46\% overall) NL-specific parameter-related bugs are due to setting incorrect
sequence length, which can affect the robustness of the system.
The model creators commonly set up the sequence or max length. This parameter limits
the length of the sentence that can be processed at a time. For instance, \bert
and \roberta have a maximum allowable sequence length of 512. While this
constraint can certainly help the model perform better, it can have adverse
effects. 
If the average length of the sentences in a dataset is more than the max length,
then the model may not learn the reference, which the adversary can utilize
to break functionalities. For instance, as shown in Figure \ref{fig:max}, the
sentence ``The man went to the store and bought a gallon of milk'' will be split
between two parts, 1) ``the man went to the store'', and 2) ``and bought a
gallon of milk'' based on the restriction imposed by the sequence length (for illustrative purpose, we take sequence length as 6).
Due to the constraint on the parameter, the meaning of the sentence is lost, and
an adversary can change a single word. For
example, if the word ``store'' is changed to ``mars'', the first
part of the sentence can still be valid. But
without the imposed constraint, two sentences will not be split, and the semantic will be preserved. 
In the same figure, the author of the post, a co-author of \bert suggested that a data split can help escape such problems.

\textbf{\textit{Implications.}}
There are works to auto-tune hyper-parameters in the 
DL~\cite{deng2013new, ilievski2017efficient} and SE~\cite{fu2017easy, arcuri2011parameter} domains. For instance,
AutoML~\cite{he2021automl} uses neural architecture search (NAS) for parameter
optimization. A similar architecture could be used for the NL domain. While
optimizing the parameters, besides accuracy, other non-functional properties should also be
taken into account. In SE, works on search-based
systems~\cite{arcuri2013parameter, agrawal2018wrong} have proposed different
techniques to identify the near-optimum value for the hyper-parameters. Such
systems could be used to tune the hyper-parameters in NL-specific software.

\subsubsection{Language.} 
\finding{Improper language specification can propagate bias in NLP pre-trained models.}
\begin{figure}[!htp]
	\includegraphics[width=1\linewidth]{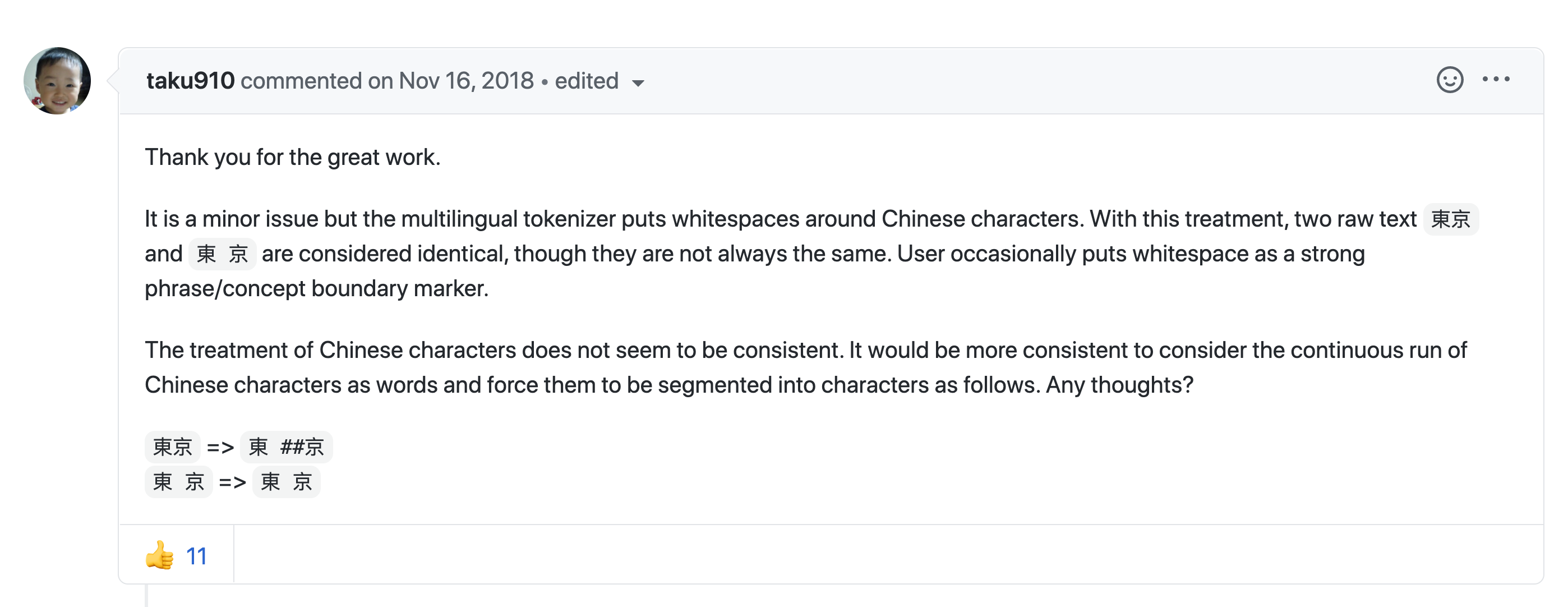}
	\caption{Natural language specification bug ~\cite{language}}
	\label{fig:language}
\end{figure}
16.67\% (3.56\% overall) of bugs related to the wrong specification of NL-parameters are due to the wrong use of NL specifications while reusing the models. For instance, Figure~\ref{fig:language} shows an issue that a developer faces while reusing the multilingual \bert for the Chinese language. Since there is no whitespace needed to separate words in Chinese, the multilingual pre-trained model often confuses while tokenizing. The pre-trained model is trained on the English dataset, and the same has been used to create the embedding of the Chinese language. However, the issue occurs due to the mismatch between the specifications of the two natural languages. While these issues can cause wrong translation by introducing whitespace, often it can invoke gender bias, too. For example, the words in English are not gender-specific, whereas
languages like Spanish, French, Hindi, etc., have gender-specific words. For
instance, in Spanish, the word ``doctor'' is either ``doctor'' (male) or
``doctore'' (female) based on the actor in the sentence.
If a model is built to translate sentences from English to Spanish, then when
the gender-specific words are encountered, the model will depend on the context
of the surrounding words. For instance, when ``My friend is a doctor''
is translated to Spanish, it can either be considered as, 1) ``Mi amigo es doctore'' (female), or 2) ``Mi amigo es doctor''
(male)~\cite{johnson2020scalable}. However, due to the presence of gender bias in
the English dataset, the system
predicts the second option or the male version. 

\textbf{\textit{Implications.}} Recently, ML models are accused of propagating bias or unfairness in the prediction \cite{galhotra2017fairness, aggarwal2019black, udeshi2018automated}. The NLP models also exhibit such bias because of their nature and reuse scenario. 
Based on the above finding, we think a formal specification should be incorporated while building such models. This is akin to introducing contracts in software development. If we compare the source language as the subtype of the target language, then a specification could be built around the multilingual translation that will validate the operation beforehand. The source language should have all the target language characteristics and more. For instance, if one of the meta-variable on which the contract can be invoked is gender, then the value corresponding to the English language will be gender-neutral, whereas it will be gender-specific for Spanish. Having such specifications can warn the developers about information loss.

\finding{Bugs can occur due to incorrect semantic preservation.}
\subsubsection{Tokenization.} 
24.07\% (5.15\% overall) of parameter-related bugs occur because of wrong tokenization,
mostly ending up crashing (68.2\%) the system. In NLP-based systems, tokenization has
been done to preserve the semantics of the input dataset and dismantle the input
dataset into smaller units, where the units can be word, sub-word, or a
character. 
\begin{figure}
	\includegraphics[width=1\linewidth]{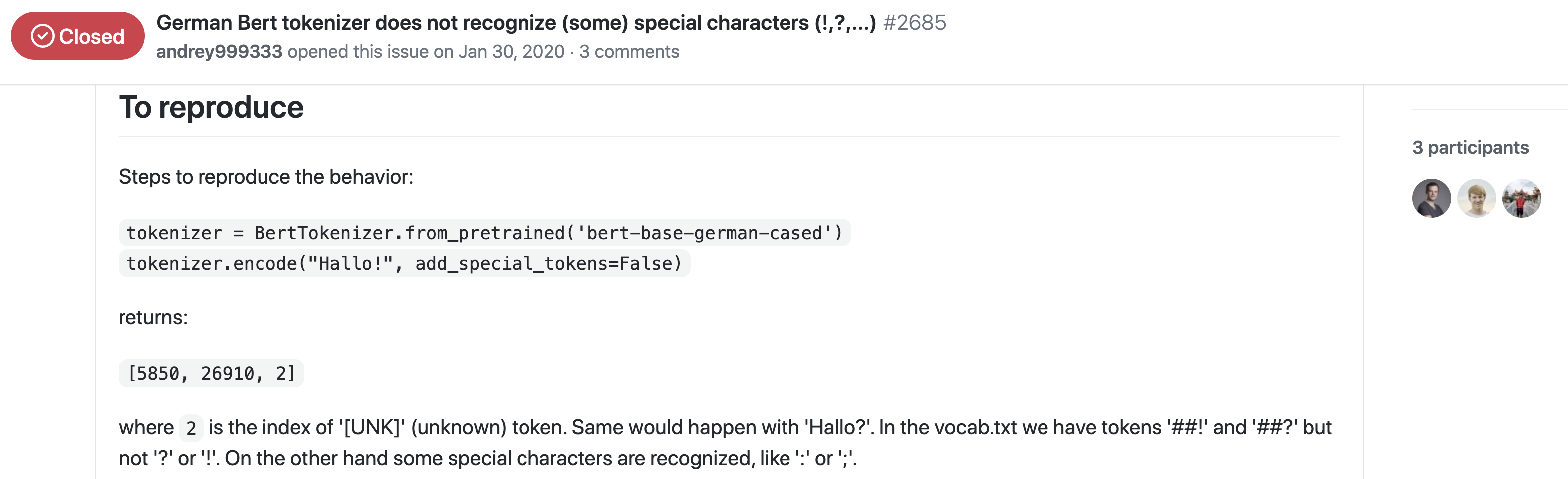}
	\caption{Example of tokenization-related bug~\cite{german}}
	\label{fig:tokenizer2}
\end{figure}
For instance, we show an example in Figure~\ref{fig:tokenizer2}, where a developer builds a model to translate German sentences into
English sentences. A model trained with the English language has been reused. If an unseen word occurs in the German language, wrong token
representation can lose the semantics of the input language. For instance,
``Hallo'' in German means ``Hi''. If the pre-trained model is reused to translate
``Hallo'', then the output of the tokenization step could be ``Hall'',
``\#\#o''. Surprisingly, there is a word ``Hall'' in the English vocabulary. In this scenario, the semantics of the word ``Hallo'' is not preserved.



\textbf{\textit{Implications.}}
In SE and PL, vast works~\cite{ouni2012search, chlipala2007certified, zhao2012formalizing} has been done to preserve the
semantics of the code during code translation. Also, a proof checking-based
approach can be implemented to validate the semantics of the source and the
target language. 
Recently, \citeauthor{lagouvardos2020static}~\cite{lagouvardos2020static} have proposed a Tensor type to validate the tensor-based operations in DL.
Such type of system can help in the NLP domain to formally prove the preservation and progress.

The rest of the (20.5\%) bugs in this category are caused due to the incorrect choice of the vocabulary length and wrong sentences given in the dataset.

\section{Frequent Impacts}
\label{sec:rq3}
In this RQ, we study the common effects of the bugs while reusing NLP models. We found that the categories denoted by the prior work~\cite{islam2019comprehensive} suffice to represent the bugs in NLP. However, the distributions are not the same. For example, incorrect functionality (DL: 9.18\%) and memory out of bound (DL: 0.82\%) errors are more prevalent in NLP (16.0
\% and 6.4\%) than traditional deep learning software. The percentages of bugs that abruptly halt the program (68.3\%) and cause bad performance (8.1\%) are less frequent in NLP. First, we discuss each type of impact and highlight the variabilities.
\begin{figure}
	\centering
	\includegraphics[width=.95\linewidth]{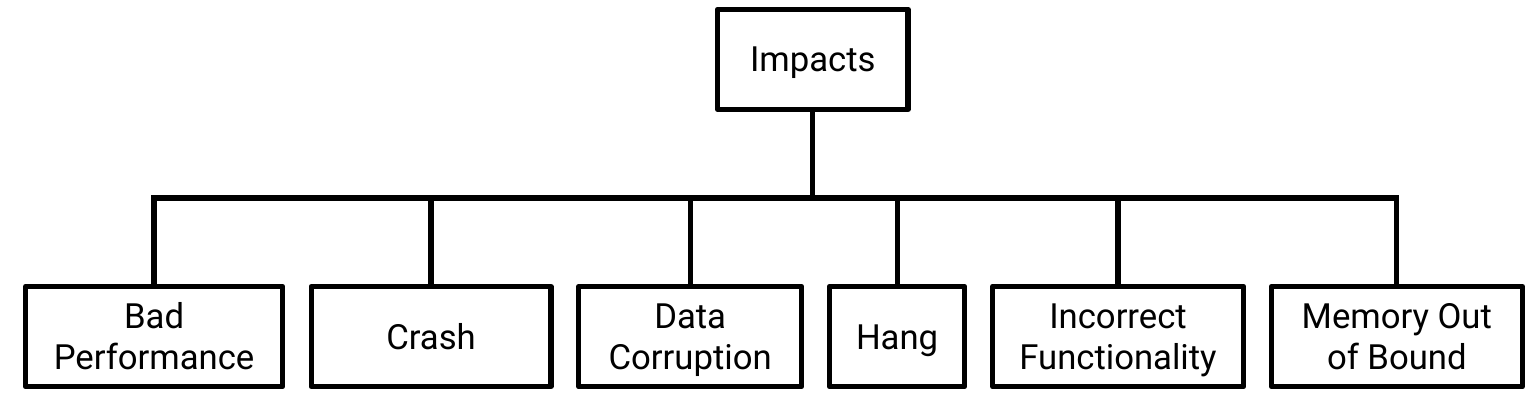}
	\caption{Classification of impacts}
	\label{fig:impactTaxo}
\end{figure}

\textbf{Bad Performance.} 
Developer often finds that the model is not performing adequately. Such bugs are classified into this category.

\textbf{Crash.} When a program exits with an error, the effect of the bugs
is classified under this category. 

\textbf{Data Corruption.} If the output data has been changed unexpectedly, then data corruption occurs. 

\textbf{Hang.} If a program does not return output for a
stipulated time and keeps running, then it generally enters the
hanging state.

\textbf{Incorrect Functionality.} Often, due to a bug in the NLP system, the output of the program is different from the expected behavior.

\textbf{Memory Out of Bound.} Often, a program in NLP halts due to the
unavailability of the memory.
%

\begin{figure}[htp]
	\centering
	\includegraphics[, trim={2cm 8cm 0cm 6cm},width=0.6\linewidth]{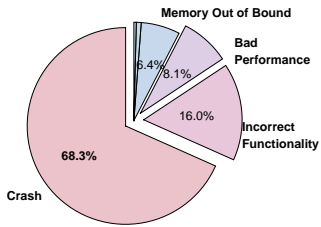}
	\caption{Distribution of impacts (labels $\le$0.7\% are
		hidden)
	}
	\label{fig:impact}
\end{figure}


\finding{Reusing a pre-trained model helps to reduce performance related bugs.}
Compared to the traditional DL software, developers prefer to reuse the pre-trained model in the NLP domain. One of the prominent reasons is to reuse the knowledge from the huge corpus utilized to train these models. The culture of reusing instead of building every solution has helped to reduce the performance-related issues significantly (in DL: 13.8\%, in NLP: 8.1\%). 
Out of all the root causes, updating the models (47.5\%) and changing the NL-specific parameters (31.25\%) are the most prevalent reasons for bad performance.

\section{Related Works}
\label{sec:related}

There is a vast body of works in DL bug study~\cite{islam2019comprehensive, islam2020repairing, thung2012empirical, zhang2018empirical, garcia2020comprehensive, humbatova2020taxonomy, wardat21deeplocalize, zhang2021autotrainer, nikanjam2021automatic, schoop2021umlaut, chakraborty2021does, liu2021reproducibility}. Here, we discuss the closest works.

Thung~\etal~\cite{thung2012empirical} have studied three machine learning systems, Apache Mahout, Lucene, and OpenNLP. Bug-type and their severity have
been identified for these systems. Though this dataset has a system
related to NLP, they did not focus on NLP model reusability.

\citeauthor{chen2021empirical}~\cite{chen2021empirical} have studied the deployment faults of DL-based mobile applications using 304 deployment faults from \sof and \gh. They have provided a taxonomy with 23 fault categories and corresponding fix strategies. Whereas we study the bugs while reusing the pre-trained NLP models.

\citeauthor{zhang2018empirical}~\cite{zhang2018empirical} have studied the software built using the \tensor
library. This work is done on 175 bugs found in \sof posts, and \gh
commits. These bugs are classified into symptoms and causes. Though this
work has charted the course in studying DL bugs, the bugs
found in traditional DL libraries are significantly different from
those found while reusing the NLP models. 

Islam~\etal~\cite{islam2019comprehensive} have studied bugs from five different
DL libraries using 2716 \sof posts and 500 \gh commit. They 
found 415 bugs from \sof posts and 555 from \gh commits. This work
classified the bugs into the root cause, bug type, and effect. Furthermore,
they studied bugs found in different pipeline stages and the presence of
different anti-patterns.
Our classification scheme has been adapted from this work. However, we found that the
NLP-model bugs are significantly different. 

Humbatova~\etal~\cite{humbatova2020taxonomy} have studied 1059 \sof and \gh
codes and developed a taxonomy of the faults seen in these posts and commits. 
This study also focused on traditional DL and did not consider the reusability of models in the NLP domain.

Chakraborty~\cite{chakraborty2021does} has studied 80 \sof posts to determine the type of bugs occurrence when reusing, particularly BERT models. We study the reuse of the 11 popular NLP models. We mined 9,214 issues from \gh and identified 984 bugs. Also, we provide a taxonomy with bug types, root causes, and impacts. 
\section{Threats To Validity}
\label{sec:threat}

\textbf{Internal Threat.} 
The finding and the implications are drawn based on the classification scheme developed. To remove the threat regarding the quality of the classification scheme, we adapted the base scheme from prior works and added categories based on an open coding approach. The classification scheme was developed by two researchers and with rigorous discussions, which is the same way prior works~\cite{islam2019comprehensive, islam2020repairing, zhang2018empirical} have developed their classification schemes. Also, to remove the researcher's bias, we compute the Cohen's Kappa coefficient to measure the agreement. Only when the perfect agreement was achieved, the raters labeled the post individually. Moreover, any discrepancies were resolved over a discussion.


\noindent\textbf{External Threat.}
The quality of the issues mined can be an external threat. To mitigate, we selected the most popular NLP models from a vastly used framework (Huggingface Transformer). Then, instead of selecting the issues that are related to bugs in \textit{Huggingface Transformer}, we mine the bugs found while using these NLP models. Then, we removed \gh repositories that are not well-maintained (number of issues $\le$50) and selected the popular ones (based on star count). We mined the bug-related issues using keywords proposed by the prior work~\cite{garcia2020comprehensive} and labeling done by the maintainer of the repositories. After mining such issues, raters validated whether the issue was related to a bug or not by manual verification. 
\section{Conclusion}
\label{sec:conclusion}

With the increase in the popularity of the NLP domain, developers are facing several bugs while reusing the pre-trained models. In this study, we mined 9214 issues from 11 repositories of well-known pre-trained models and identified 984 bugs. Then, we manually studied them to understand the common bug type, root causes, and impacts. We built a classification scheme based on an open coding scheme. We determined that the root causes of the bugs are significantly different from the bugs found in traditional deep learning. Specifically, we found that large models are bug-prone and cause memory-related issues. We identified that a parameter tuning and validation-based approach could be helpful to increase the robustness of such systems. We also identified bugs related to the propagation of input bias to the output, loss of semantic preservation in the system, etc. Lastly, we suggest different ways to prevent such issues (i.e., verify the system to ensure semantic preservation, etc.). Our findings can help guide both the NL and SE practitioners and researchers through the most prevalent problems of reusing these models and help build automated repairing approaches to address the same.

%

\balance
\bibliographystyle{ACM-Reference-Format}
\bibliography{refs}


\end{document}